\documentclass{article}
\usepackage{epsfig}
\usepackage{amsmath}
\usepackage{amsfonts}
\usepackage{graphicx}
\usepackage[german, english]{babel}
\usepackage{a4wide}
\usepackage{amsmath}
\usepackage{amssymb}
\usepackage{ifthen}
\usepackage{epsfig}
\newcounter{fig}   \newcommand{\lbfig}[1]{\refstepcounter{fig}
\label{#1} }

\newcommand{\bea}{\begin{eqnarray}}
\newcommand{\eea}{\end{eqnarray}}
\newcommand{\be}{\begin{equation}}
\newcommand{\ee}{\end{equation}}

\newcommand{\re}[1]{(\ref{#1})}

\begin{document}

\title{Gauged Hopfions}

\author{
{\large Ya. Shnir}$^{\dagger \star}$
and {\large G. Zhilin}$^{\star}$ \\ \\
\\ $^{\dagger}${\small BLTP, JINR, Dubna, Russia}
\\ $^{\star}${\small Department of Theoretical Physics and Astrophysics}\\
{\small Belarusian State University, Minsk 220004, Belarus}
} \maketitle

\begin{abstract}
We discuss the $U(1)$ gauged version  of the 3+1 dimensional
Faddeev-Skyrme model supplemented by the Maxwell term.
We show that there exist axially symmetric static solutions coupled to the
non-integer toroidal flux of
magnetic field, which revert to the usual Hopfions ${\cal A}_{m,n}$ of lower degrees $Q=mn$
in the limit of the gauge coupling constant vanishing.
The masses of the static gauged Hopfions are found to be less than the corresponding masses
of the usual ungauged solitons ${\cal A}_{1,1}$ and ${\cal A}_{2,1}$ respectively, they become lighter as gauge coupling
increases.
The dependence of the solutions on the gauge coupling is investigated.
We find that in the strong coupling regime the gauged Hopfion carries two
magnetic fluxes, which are
quantized in units of $2\pi$, carrying $n$ and $m$ quanta respectively.
The first flux encircles the position curve and the second one is directed along the symmetry axis.
Effective quantization of the field in the gauge sector
may allow us to reconsider the usual arguments concerning the lower topological bound in
the  Faddeev-Skyrme-Maxwell model.
\end{abstract}


\section{Introduction}

Spatially localized finite energy particle-like soliton solutions play a prominent role
in a wide variety of non-linear physical systems, from modern cosmology and quantum field theory
to condensed matter physics (for a general review see e.g. \cite{Manton-Sutcliffe}).

There is a simple example of topological solitons given by the class of non-linear scalar
models of the Skyrme family. In $d=3+1$ dimensions it includes the original
Skyrme model \cite{Skyrme:1961vq} and the Faddeev-Skyrme model \cite{Faddeev}, in $d=2+1$
there is a simplified  baby Skyrme model \cite{BB,Bsk} which resembles the basic properties of the
genuine Skyrme model in many aspects. A unifying
feature of all these models is that their structure is identical, the corresponding Lagrangians
include the usual sigma model term,
the Skyrme term, which is quartic in derivatives of the field, and the potential term which does not
contain the derivatives. Note that the latter term is optional
in $d=3+1$, however it is obligatory to stabilise the soliton solutions
of the low-dimensional baby-Skyrme model \cite{Derrick}.

A peculiar feature of all these models is that the corresponding
soliton solutions, both Skyrmions and Hopfions, do not saturate the
topological lower bound. In order to attain it
and get a relation\footnote{This relation is linear for Skyrmions, however for the Hopfions
the corresponding Vakulenko-Kapitanski bound is $E=c Q^{3/4}$
where $c=(3/16)^{3/8}$ \cite{VK}. } between the masses of the solitons and
their topological charges $Q$, one has to modify the model, for
example eliminate quadratic in derivatives term
\cite{Adam:2010fg,Foster:2010zb} or extend the model by coupling
of the Skyrmions to an infinite tower of vector mesons
\cite{Sutcliffe:2011ig}.

A physically natural extension of the Skyrme model is related with possibility of
gauging of the global symmetry group. Originally, the Abelian
gauged Skyrme model was proposed to model the monopole catalysis of the proton decay
\cite{Callan:1983nx}, the axially-symmetric gauged Skyrmions were considered in  \cite{Piette:1997ny}.
Similar analysis of the gauged baby Skyrmions \cite{Gladikowski:1995sc} reveal very interesting features of the
corresponding solitons: they carry a magnetic flux which is not topologically quantized.
Gauge versions of some systems, which resemble a modified Faddeev-Skyrme model,
were also considered  in the papers \cite{Babaev:2001zy,Protogenov:2002bt,Radu:2005jp,Jaykka:2011zz}.
However, to the best of our knowledge, the analysis of the fully coupled Faddeev-Skyrme-Maxwell
system has not yet been done.

Another peculiarity of the soliton solutions of the Skyrme family is that
they can be constructed only numerically, one has to apply rather complicated numerical
methods which need a serious amount of computation power.
This task becomes particularly complicated in the case of the Hopfions
in the Faddeev-Skyrme model which are string-like configurations
classified by the linking number, the first Hopf map $S^3 \to S^2$.

The Hopfions have been intensively studied over
recent years \cite{Gladikowski:1996mb,Battye1998,Sutcliffe:2007ui,Hietarinta2000}.
These solitons have a number of physical applications,
for example, in study of Bose-Einstein condensates \cite{Kawaguchi:2008xi}, nonlinear optics \cite{Dennis2010}
and non-conventional superconductivity \cite{Babaev:2008zd}.

It was shown that whereas the minimal energy solitons  of the Faddeev-Skyrme model
of degree $Q=1,2$ are axially symmetric, the higher degree solutions should be not just
closed  flux-tubes of the fields but knotted field configurations.
Note that the solitons of the model possess both rotational and
internal rotational (or isorotational) degrees of freedom, such a rotation
might seriously affect the structure of the Hopfions \cite{BattyMareike,JHSS}.

The soliton solutions of the Faddeev-Skyrme model are invariant with respect to the global $SO(2)$ symmetry.
Furthermore, for axially symmetric configurations the rotations about the third axis in space and in isospace
are identical. Therefore, by analogy with the similar situation in the planar baby Skyrme model \cite{Gladikowski:1995sc},
we can couple the usual Maxwell electrodynamics to the Faddeev-Skyrme model by gauging
this symmetry. Note that in such a theory, unlike the low-dimensional baby Skyrme model \cite{Schroers:1995he},
the Maxwell term alone probably cannot be used as a substitute for the  Skyrme term in order to stabilize the Hopfions.

In this Letter we discuss the topologically stable static
soliton solutions of the full coupled gauged Faddeev-Skyrme-Maxwell system which
carry two magnetic fluxes.
We study numerically the dependency of the shape of these gauged Hopfions, their masses
and magnetic fluxes on the gauge coupling constant, both in perturbative limit and in the strong coupling
limit. Since the consistent consideration of the solitons with higher
Hopf charges is related with complicated task of full numerical simulations in 3d \cite{Sutcliffe:2007ui},
we restrict our consideration to the case of the static axially symmetric unlinked Hopfions
${\cal A}_{1,1}$ and ${\cal A}_{2,1}$ of charges $Q=1,2$, respectively.
We find that the gauged Hopfion carries a toroidal magnetic flux which, in the strong coupling regime is effectively reduced
to two magnetic fluxes, one of which encircles the position curve of the Hopfion and the second one is directed along the
symmetry axis. Further, we demonstrated that both fluxes are quantized
in units of $2\pi$, carrying  $n$ and  $m$ quanta respectively.

More systematic investigation of
the gauged Hopfions for larger values of $Q$ and with non-vanishing electric field will be presented elsewhere.

\section{The model}
A gauged version of the Faddeev-Skyrme model can be constructed if we
take into account global $SO(2)$ invariance of the $3+1$ dimensional Lagrangian
\be
\label{model}
{\cal L}_{FS} = \frac{1}{32\pi^2\sqrt 2}\left(\partial_\mu \phi^a \partial^\mu \phi^a -
\frac{\kappa}{2}(\varepsilon_{abc}\phi^a\partial_\mu \phi^b\partial_\nu \phi^c)^2 - \mu^2 [1-(\phi^3)^2] \right)
\ee
where $\kappa$ is the dimensional coupling constant and
a triplet of scalar real fields $\phi^a = (\phi^1, \phi^2,\phi^3)$ satisfy the
constraint $\phi^a \cdot \phi^a=1$. An additional potential term $V=\mu^2 [1-(\phi^3)^2]$
breaks the global $SO(3)$ symmetry of the model. Note that if the model \re{model} becomes
restricted to the $xy$ plane, it corresponds to the double vacuum baby Skyrme model \cite{Weidig:1998ii}.

Under the scaling transformations of the domain space $x \to \lambda x$ the sigma model term scales as
$\lambda$, the Skyrme term scales as $\lambda^{-1}$ and the potential term scales as $\lambda^3$. Hence,
even if the latter term is absent, the existence of static solitons of the model \re{model}
is allowed by the Derrick theorem.

Topological restriction on the field $\phi^a$ is that it approaches its
vacuum value at spacial boundary, i.e. $\phi^a_\infty = (0,0,1)$. This allows a one-point compactification of
the domain space $\mathbb{R}^3$ to $S^3$ and
the field of the finite energy solutions of the
model, the Hopfions, is a map $\phi^a:\mathbb{R}^3 \to S^2$ which belongs
to an equivalence class characterized by the homotopy group $\pi_3(S^2)=\mathbb{Z}$. Explicitly, the
Hopf invariant is defined non-locally as
\be
\label{charge}
Q=\frac{1}{16\pi^2}\int\limits_{\mathbb{R}^3}\varepsilon_{ijk}{\cal F}_{ij} {\cal A}_k
\ee
where ${\cal F}_{ij}=\varepsilon_{abc}\phi^a\partial_i \phi^b\partial_j \phi^c$ and one-form ${\cal A} = {\cal A}_k dx^k$ is
defined via ${\cal F}=d {\cal A}$, i.e the two-form ${\cal F}$ is closed, $d{\cal F}=0$.

For the lowest values of the corresponding Hopf charge $Q=1,2$ the simplest soliton solutions can be constructed using the
axially symmetric ansatz \cite{Gladikowski:1996mb} written in terms of two
functions $f=f(r,\theta)$ and $g=g(r,\theta)$ which depend on the radial variable $r$ and the polar angle $\theta$:
\begin{eqnarray}
\label{ansatz}
\phi^1&=&\sin f(r,\theta) \cos (m\varphi -n g(r,\theta)),\nonumber\\
\phi^2&=&\sin f(r,\theta) \sin (m\varphi -n g(r,\theta)), \nonumber\\
\phi^3&=&\cos f(r,\theta) \, ,
\end{eqnarray}
where  $n,m \in \mathbb{Z}$.
An axially-symmetric configuration of
this type is commonly referred to as  ${\cal A}_{m,n}$, where the first subscript
corresponds to the number of twists along the loop
and the second label is the usual $O(3)$ sigma model winding number associated with the
map $S^2 \to S^2$. The Hopf invariant of this configuration is $Q = mn$.

The unbroken global symmetry of the configurations with
respect to the rotations around the third axis allows us to rotate
the components of the axially-symmetric configuration as
$(\phi_1 + i\phi_2) \mapsto (\phi_1 + i\phi_2)e^{i\alpha}$, where $\alpha$ is the angle
of rotation. Thus, we can gauge this subgroup by a $U(1)$ gauge field $A_\mu$ defining the covariant
derivative as (cf \cite{Gladikowski:1996mb,Schroers:1995he,Adam:2012pm})
\be
\label{cov-der}
D_\mu \phi^a= \partial_\mu \phi^a + g A_\mu \varepsilon_{abc} \phi^b \phi^c_\infty \, .
\ee
where $g$ is the gauge coupling constant.

Note that the field configuration has finite energy if $D_\mu \phi^a \to 0$ as $r \to \infty$. Hence,
on the spacial asymptotic,
the field of the gauged Hopfion must lie in an orbit of the gauge group,
unless the global symmetry is explicitly broken by the potential term. In other words
this condition generically does not imply the field $\phi^a$ necessarily tends to a constant
on the spacial asymptotic.

The total Lagrangian of the gauged Faddeev-Skyrme-Maxwell model can be written as
\be
\label{lag}
{\cal L} = \frac{1}{32\pi^2\sqrt 2}\left(-\frac{1}{4}F_{\mu\nu}F^{\mu\nu}+D_\mu \phi^a D^\mu \phi^a -
\frac{\kappa}{2}(\varepsilon_{abc}\phi^a D_\mu \phi^b D_\nu \phi^c)^2 - \mu^2 [1-(\phi^3)^2] \right)
\ee
where we introduced the usual Maxwell term and the field strength tensor is
$F_{\mu\nu}=\partial_\mu A_\nu -\partial_\nu A_\mu$. Here we suppose that the topological charge of the configuration
is defined as usual by \re{charge}.

Note that  the integrated Maxwell term transforms as $\lambda^{-1}$ under the scaling transformations $x \to \lambda x$,
i.e. it has the same scaling properties as the Skyrme term. Setting $\lambda = \sqrt \kappa$ allows us to
rescale the Skyrme coupling constant to $\kappa=1$.

The question about possible existence of soliton solutions of the gauged model \re{lag} was briefly discussed
recently in review \cite{Radu:2008pp}. It was pointed out that, in some sense,
the Maxwell term looks similar to the Skyrme term, thus one can consider an additional
linking between the genuine $U(1)$ gauge field $A_\mu$ and the "potential" ${\cal A}_{\mu}$ which appears in the definition
of the Hopf charge \re{charge} \cite{Radu:2008pp,Protogenov:2002bt}. However it may result in instability of the configuration
since it seems to be there is no topological restrictions on the Maxwell field
and there is no lower energy bound in the generalized
Vakulenko-Kapitanski relation \cite{Protogenov:2002bt}. However, as we will see below, there still is an
effective quantization of the field fluxes in the gauge sector which may
allow us to evade a possible collapse of the configuration.

Hereafter, we restrict the consideration to the original model without the potential term,
so we set $\mu=0$. Then in normalized units of energy, in which $E \to E/(32\pi^2)\sqrt 2$,
the static energy of the gauged Hopfion is defined by the functional
\be
\label{E}
E = \int\limits_{\mathbb{R}^3}\left[(D_i \phi^a)^2  +
\frac{1}{2}(\varepsilon_{abc}\phi^a D_i \phi^b D_j \phi^c)^2\right]
\ee
and the electromagnetic part of the total energy functional is the usual sum of the magnetic and electric
components:
\be
\label{E_g}
E_{em}=\frac{1}{2}\int\limits_{\mathbb{R}^3}\left[B_k^2 + E_k^2\right]
\ee
Let us consider purely magnetic field generated by the axially symmetric Maxwell potential
\be
\label{el-ansatz}
A_0 = A_r = 0;\qquad A_\theta=A_1(r,\theta);\qquad A_\phi=A_2(r,\theta) \sin \theta
\ee
represented in terms of the two functions $A_i(r,\theta)~~i=1,2$, here the gauge fixing condition is
used to exclude the radial component of the vector-potential.
Note that the
"trigonometric" parametrization \re{ansatz} is not very convenient from the point of view of numerical
calculations \cite{Acus:2012st,Shnir:2009ct}, here we used it to produce an initial configurations in the given
topological sector. However the original triplet of the scalar fields $\phi^a$
was considered as dynamical variables in the
corresponding system of the Euler-Lagrange equations.

Certainly, the $U(1)$ gauge potential obeys the usual Maxwell equation
\be
\label{maxwell}
\partial_\mu F^{\mu\nu} = j^\nu
\ee
with the current
\be
\label{current}
j_\mu = 2g\varepsilon_{abc}\phi^a D_\mu \phi^b \left( \phi^c_\infty - D_\nu \phi^c \partial^\nu \left(\phi^d \phi^d_\infty\right) \right).
\ee
The complete set of the field equations, which follows from the variation of the action of the Faddeev-Skyrme-Maxwell
model \re{lag}, can be solved when we impose
the boundary conditions.  As usually, they follow from the regularity on the symmetry axis
and symmetry requirements as well as
the condition of finiteness of the energy and the topology. In particular we have to take into account that the magnetic field
is vanishing on the spacial asymptotic.  Explicitly, we impose
\begin{equation}
\phi^{1}\biggl.\biggr|_{r \rightarrow \infty }\!\!\!\rightarrow
0\,,~~\phi^{2}\biggl.\biggr|_{r \rightarrow \infty
}\!\!\!\rightarrow 0\,,~~\phi^{3}\biggl.\biggr|_{r \rightarrow
\infty }\!\!\!\rightarrow 1\,,~~\partial_r A_1\biggl.\biggr|_{r \rightarrow \infty
}\!\!\!\rightarrow 0\,,  ~~A_2\biggl.\biggr|_{r \rightarrow \infty
}\!\!\!\rightarrow 0\,,\label{infty1}
\end{equation}%
at infinity and
\begin{equation}
\phi^{1}\biggl.\biggr|_{r \rightarrow 0 }\!\!\!\rightarrow
0\,,~~\phi^{2}\biggl.\biggr|_{r \rightarrow 0
}\!\!\!\rightarrow 0\,,~~\phi^{3}\biggl.\biggr|_{r \rightarrow
0 }\!\!\!\rightarrow 1\,,~~ A_1\biggl.\biggr|_{r \rightarrow 0
}\!\!\!\rightarrow 0\,,  ~~A_2\biggl.\biggr|_{r \rightarrow 0
}\!\!\!\rightarrow 0\,,\label{origin}
\end{equation}%
at the origin. The condition of regularity of the fields on the symmetry axis yields
\begin{equation}
\label{bound-axis}
\phi^{1}\biggl.\biggr|_{\theta \rightarrow 0,\pi }\!\!\!\rightarrow
0\,,~~\phi^{2}\biggl.\biggr|_{\theta \rightarrow 0,\pi
}\!\!\!\rightarrow 0\,,~~\phi^{3}\biggl.\biggr|_{\theta \rightarrow 0,\pi}
\!\!\!\rightarrow 1\,,~~ A_1\biggl.\biggr|_{\theta \rightarrow 0,\pi
}\!\!\!\rightarrow 0\,,  ~~A_2\biggl.\biggr|_{\theta \rightarrow 0,\pi
}\!\!\!\rightarrow 0\,,
\end{equation}%

\section{Numerical results}
The numerical calculations are mainly performed on a equidistant grid
in spherical coordinates $r$ and $\theta$, employing the compact radial coordinate $x= r/(1+r) \in
[0:1]$ and $\theta \in [0,\pi]$.
To find solutions of the Euler-Lagrange equations which follow from the Lagrangian \re{lag} and
depend parametrically on coupling constant $g$,
we implement a simple forward differencing scheme on a rectangular lattice with lattice spacing $\Delta x = 0.01$.
Typical grids used have sizes $120 \times 70$. The relative errors of the solutions are of order of $10^{-4}$
or smaller. We also introduce an additional Lagrangian multiplier to constrain the field to the surface of unit
sphere.

Each of our simulations began at $g=0$ at fixed value of $\mu$,
then we proceed by making small increments in $g$.

\begin{figure}[hbt]
\lbfig{fig:4}
\begin{center}
\includegraphics[height=.23\textheight, angle =-0]{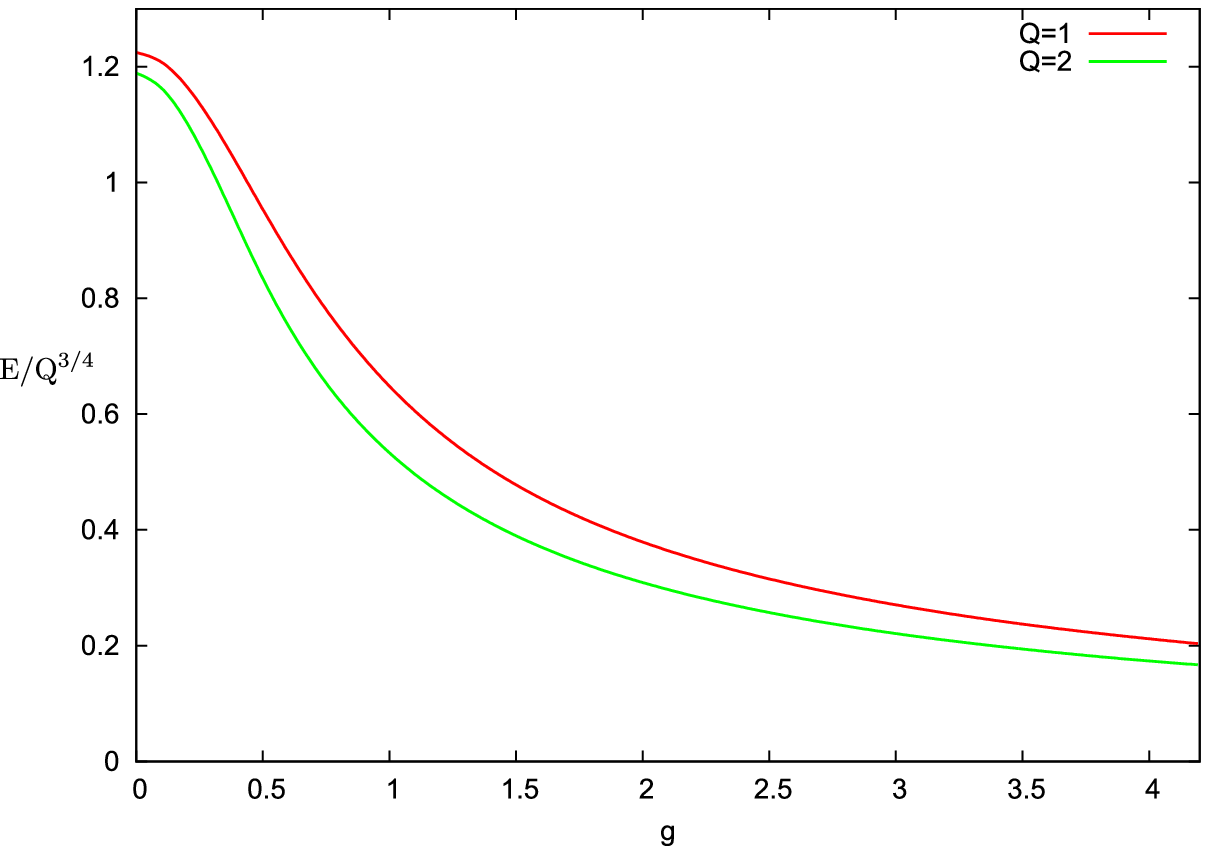}
\includegraphics[height=.23\textheight, angle =-0]{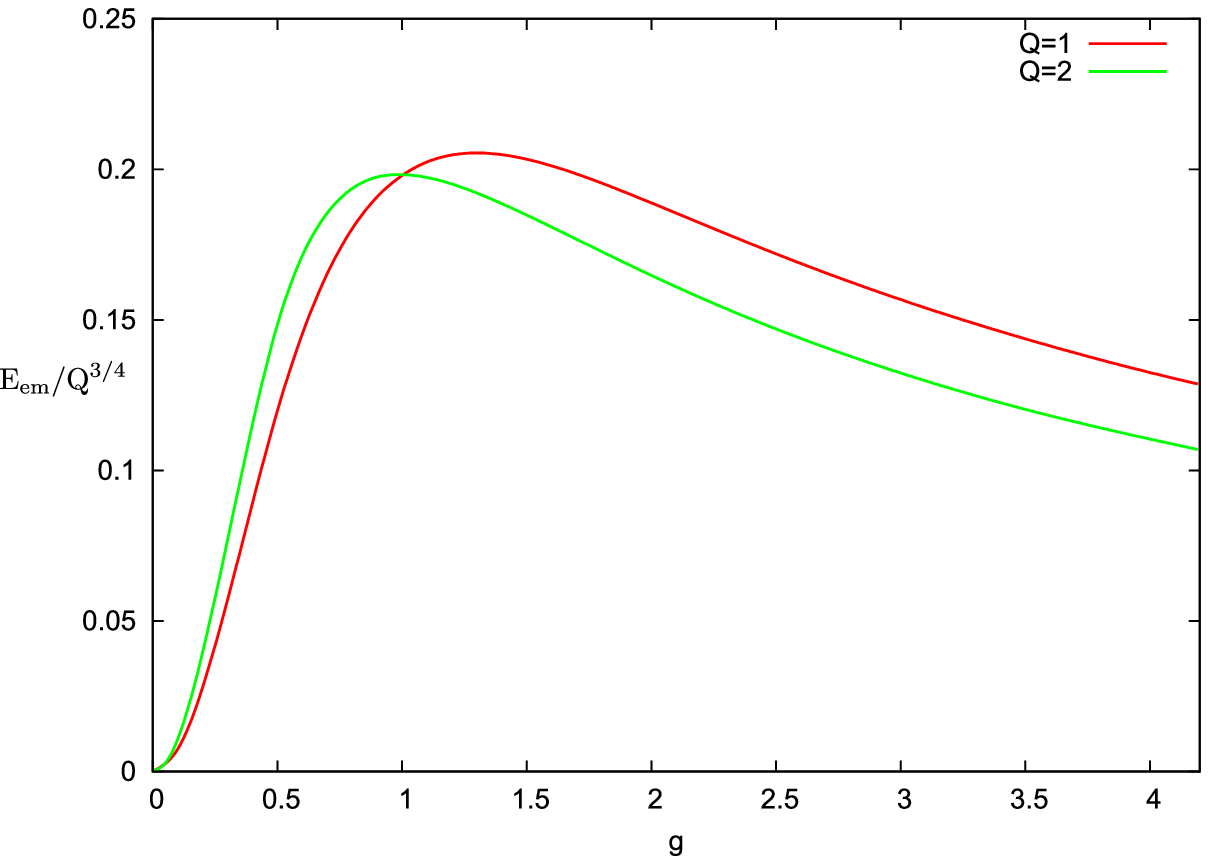}
\end{center}
\caption{\small The normalized energy $E$ of the ${\cal A}_{1,1}$ and ${\cal A}_{2,1}$  gauged Hopfions
(left plot) and the corresponding magnetic energy (right plot)
as a function of the coupling constant $g$  at $\mu=0$. }
\end{figure}

In Fig.~\ref{fig:4} we have plotted the graphs of energy of gauged Hopfion defined by the functional \re{E},
and magnetic energy as function of the gauge
coupling. Here we used the normalized units of energy and took into
account the Ward's conjecture \cite{ward} concerning the lower
energy bound, in this units it becomes  $E\geq Q^{3/4}$.

As the gauge coupling increases from zero, the energy of the gauged Hopfion decreases since the
toroidal magnetic flux is formed.  This squeezes the configuration down, as shown
in Fig.~\ref{fig:1} where we exhibited the energy density isosurfaces of the
gauged ${\cal A}_{1,1}$ and ${\cal A}_{2,1}$
Hopfions at $g=0$ and $g=2$, respectively.

Note that as the coupling remains smaller than one, the electromagnetic energy
$E_{em}$ defined by \re{E_g} is increasing, however in the strong coupling limit its contribution begins
to decrease as $g$ continue to grow, see Fig.~\ref{fig:4}, right plot.
We can understand this effect if we note that the
conventional rescaling of the potential $A_\mu \to g A_\mu $
leads to $F_{\mu\nu}^2 \to \frac{1}{g^2}F_{\mu\nu}^2$. Thus, the very large gauge coupling effectively
removes the Maxwell term
leaving the limiting configuration of gauged Hopfion coupled to a circular magnetic vortex of constant
flux. Apparently, in such a limit the strong coupling with a vortex yields an effective "mass term"
$g^2(A_1^2 + A_2^2)[1-(\phi^3)^2]/r^2$.
Unlike the usual symmetry breaking term in \re{model}, it affects both the gauge potential and
the field components $\phi^1$ and $\phi^2$ which became massive due to coupling to the gauge sector.

On the other hand, the gauge field remains massless on the symmetry axis since $\phi^1=\phi^2=0$
at $\theta=0,\pi$, due to the boundary condition \re{bound-axis} we imposed there. Furthermore, it is massless
along the position curve of the Hopfion where $\phi^3 = -1$, thus a magnetic flux should also appear there.

\begin{figure}[hbt]
\lbfig{fig:1}
\begin{center}
\includegraphics[height=.24\textheight, angle =0]{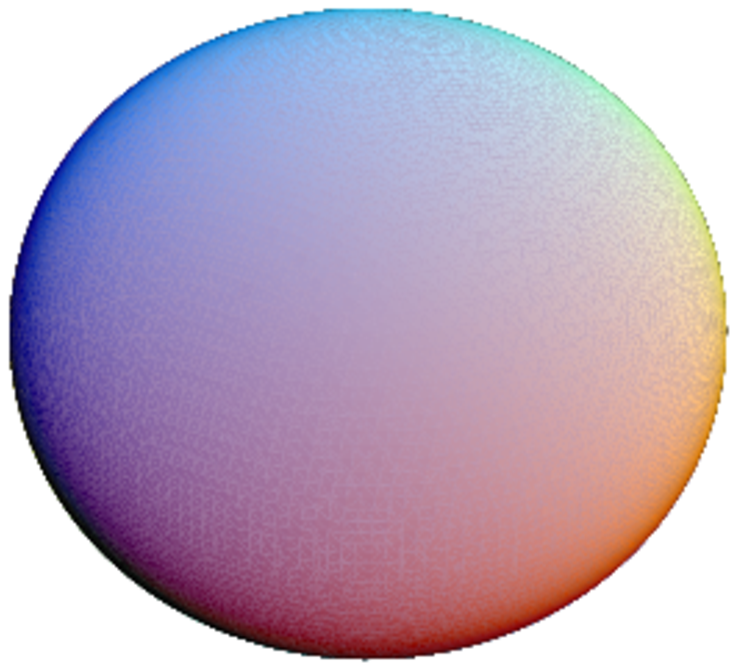}\hspace{1.5cm}
\includegraphics[height=.24\textheight, angle =0]{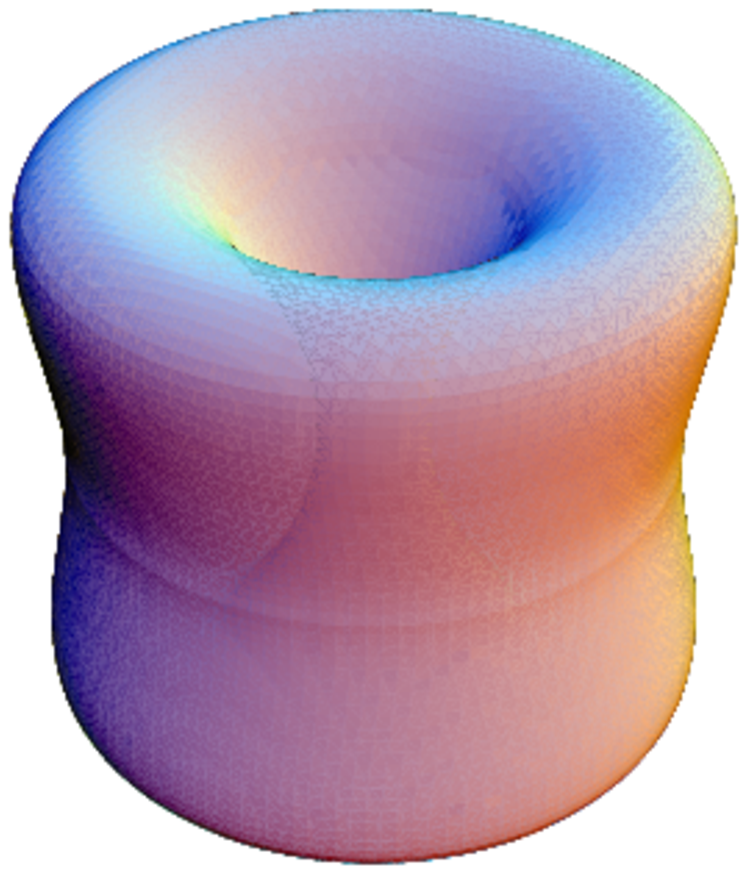}
\includegraphics[height=.24\textheight, angle =0]{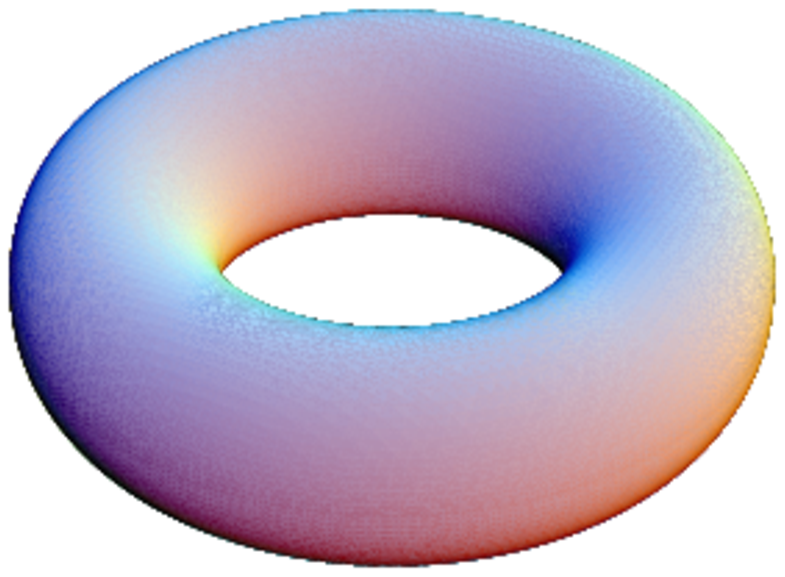}
\includegraphics[height=.24\textheight, angle =0]{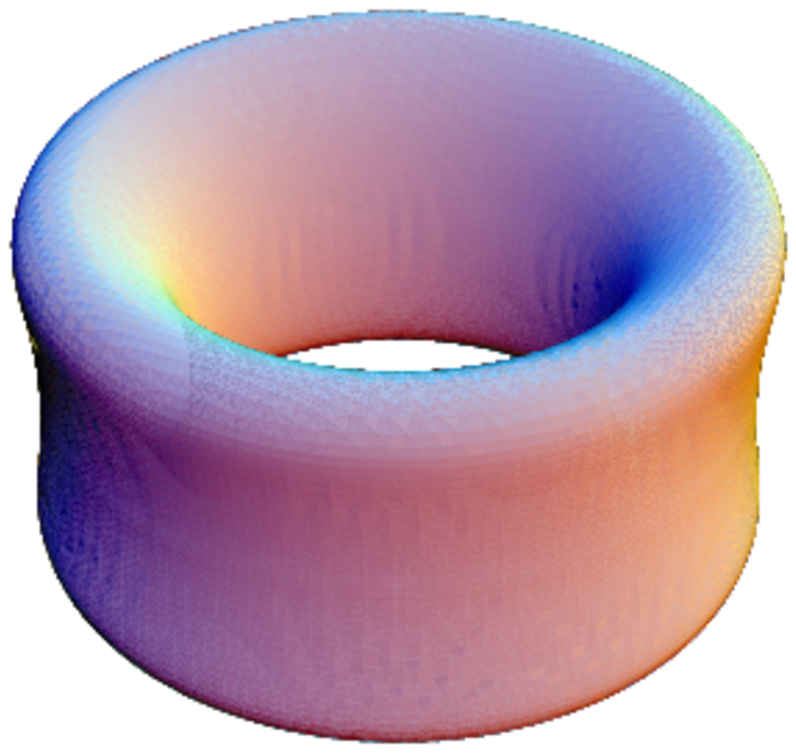}
\end{center}
\caption{\small Top row: energy density isosurfaces for gauged solitons ${\cal A}_{1,1}$  at $g=0$ (left plot) and
$g=2$ (right plot). Bottom row: energy density isosurfaces for gauged solitons ${\cal A}_{2,1}$
at $g=0$ (left plot) and $g=2$ (right plot).}
\end{figure}

Indeed, in Fig.~\ref{fig:2} we display the results of our numerical calculations of the
magnetic field of the gauged Hopfion in the $xz$ plane and in the $xy$ plane.
Evidently, in the weak coupling regime, this is a toroidal field
which encircles the position curve. The flux of such a field is not-quantized since there is
no topological reason for that. As the gauge coupling increases, the vortex is
getting smaller and the magnitude of the magnetic field increases significantly. Effectively, using the Maxwell
equation \re{maxwell}, one can set this magnetic field
into correspondence with a circular electric current $\vec j$ \cite{Shnir:2005te}.
Note that there is an interesting similarity between the magnetic flux around
the gauged Hopfion and
the magnetic field generated by the vortex-like configurations in the $SU(2)$ Yang-Mill-Higgs system
\cite{Kleihaus:2003nj,Kleihaus:2003xz,Kleihaus:2004is}

\begin{figure}[hbt]
\lbfig{fig:2}
\begin{center}
\begin{minipage}{0.45\textwidth}
\hspace{5pt}\includegraphics[height=.27\textheight, angle =0]{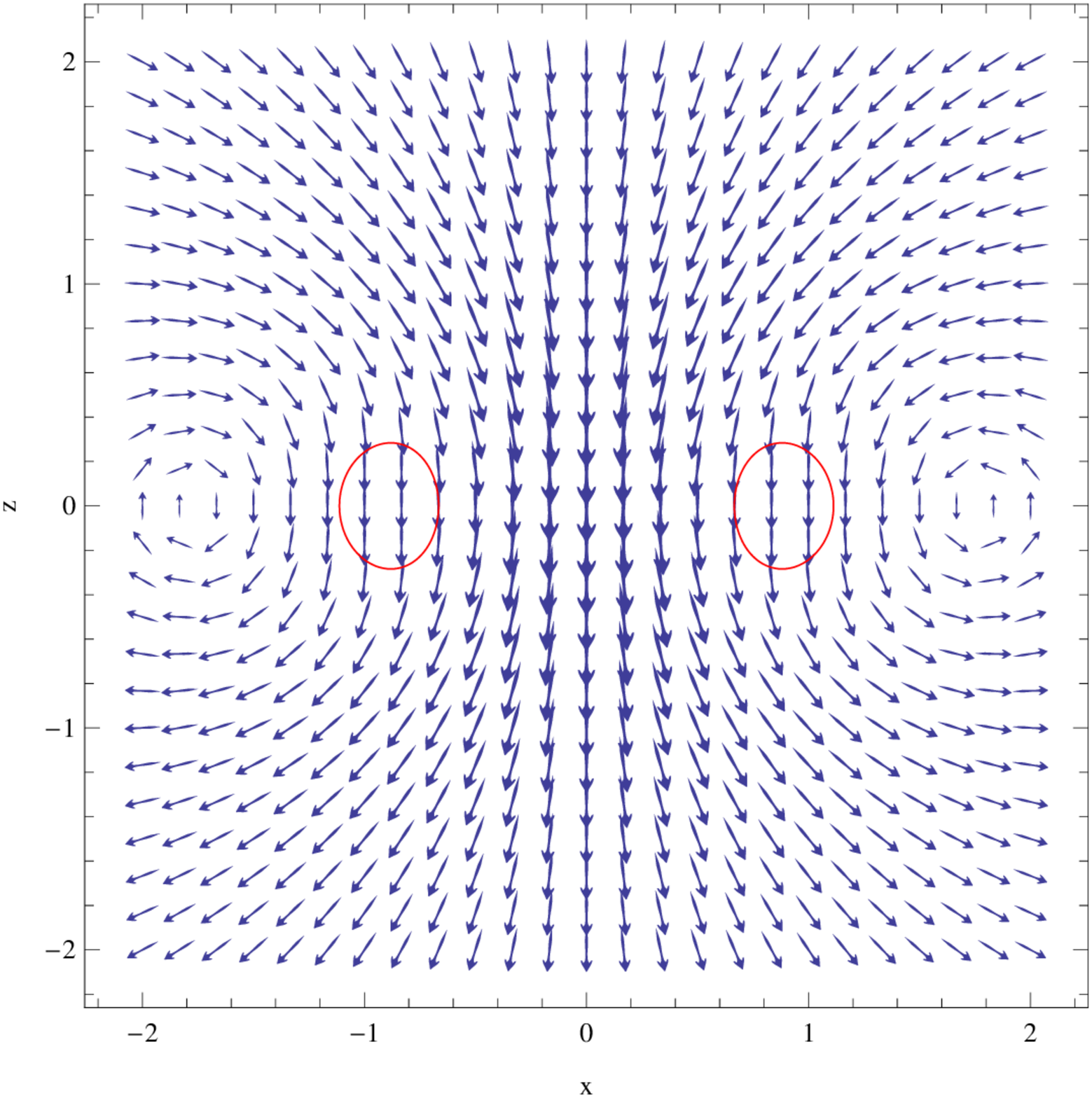}
\includegraphics[height=.27\textheight, angle =0]{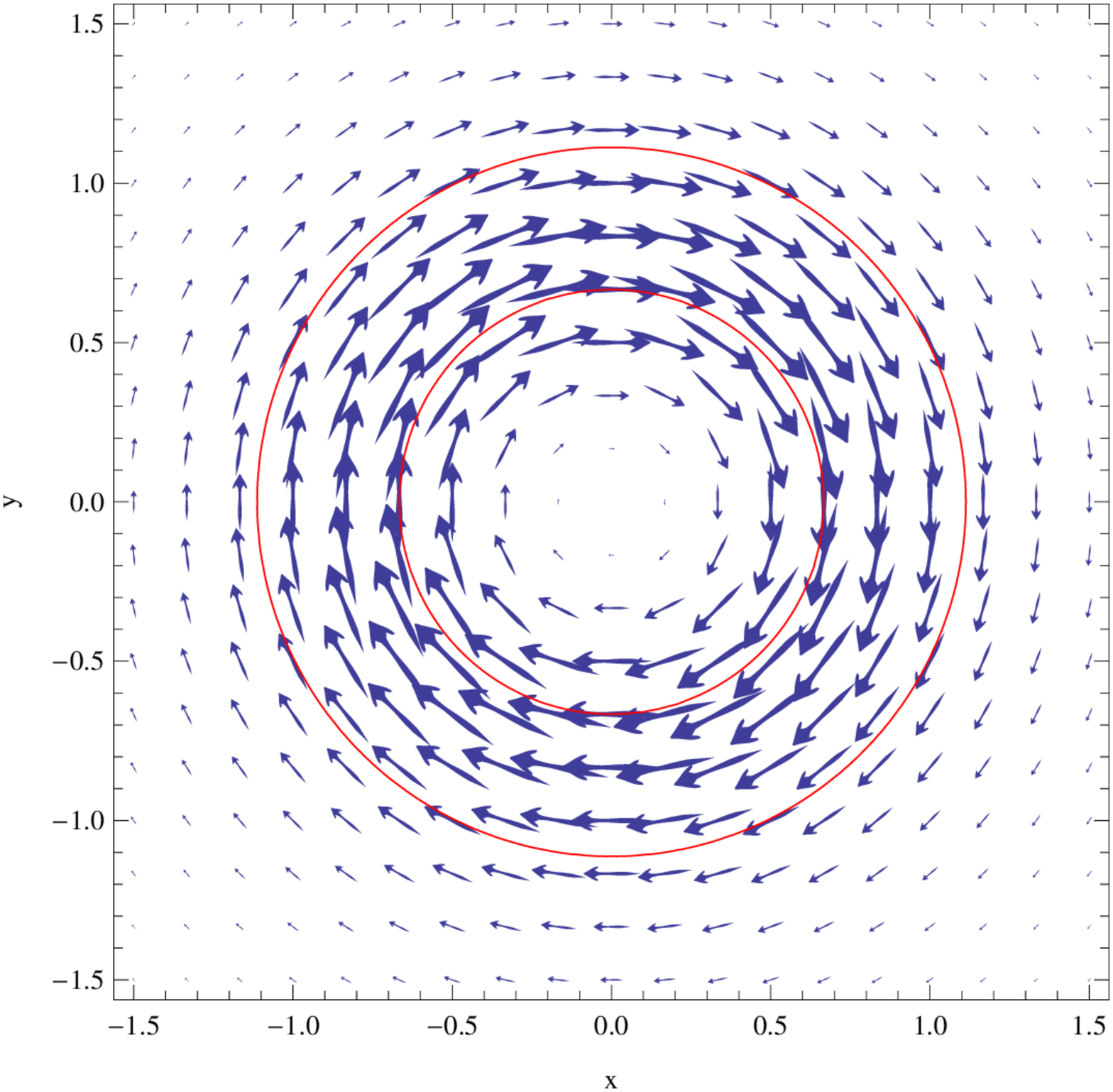}
\end{minipage}
\begin{minipage}{0.45\textwidth}
\hspace{5pt}\includegraphics[height=.27\textheight, angle =0]{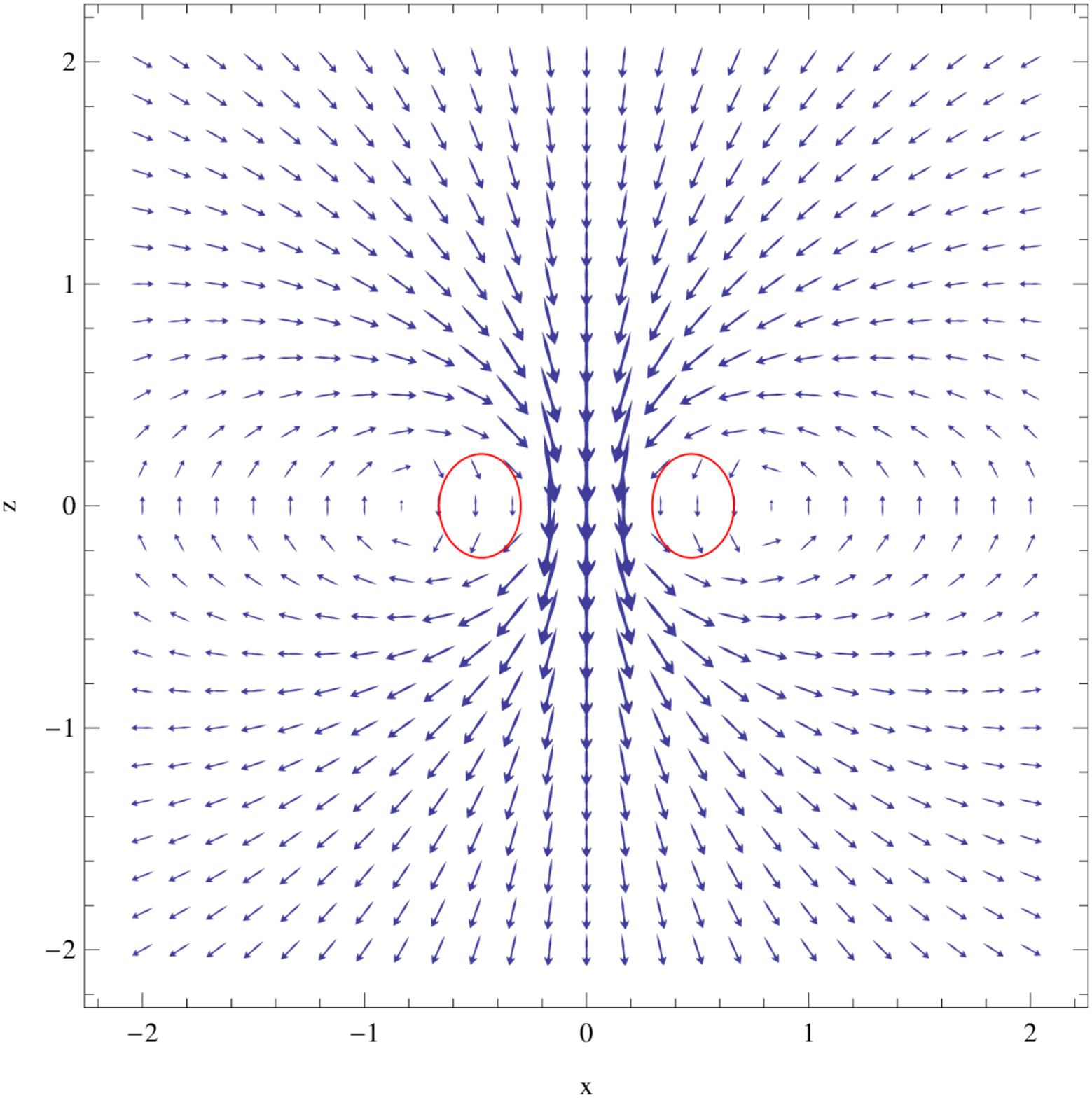}
\includegraphics[height=.27\textheight, angle =0]{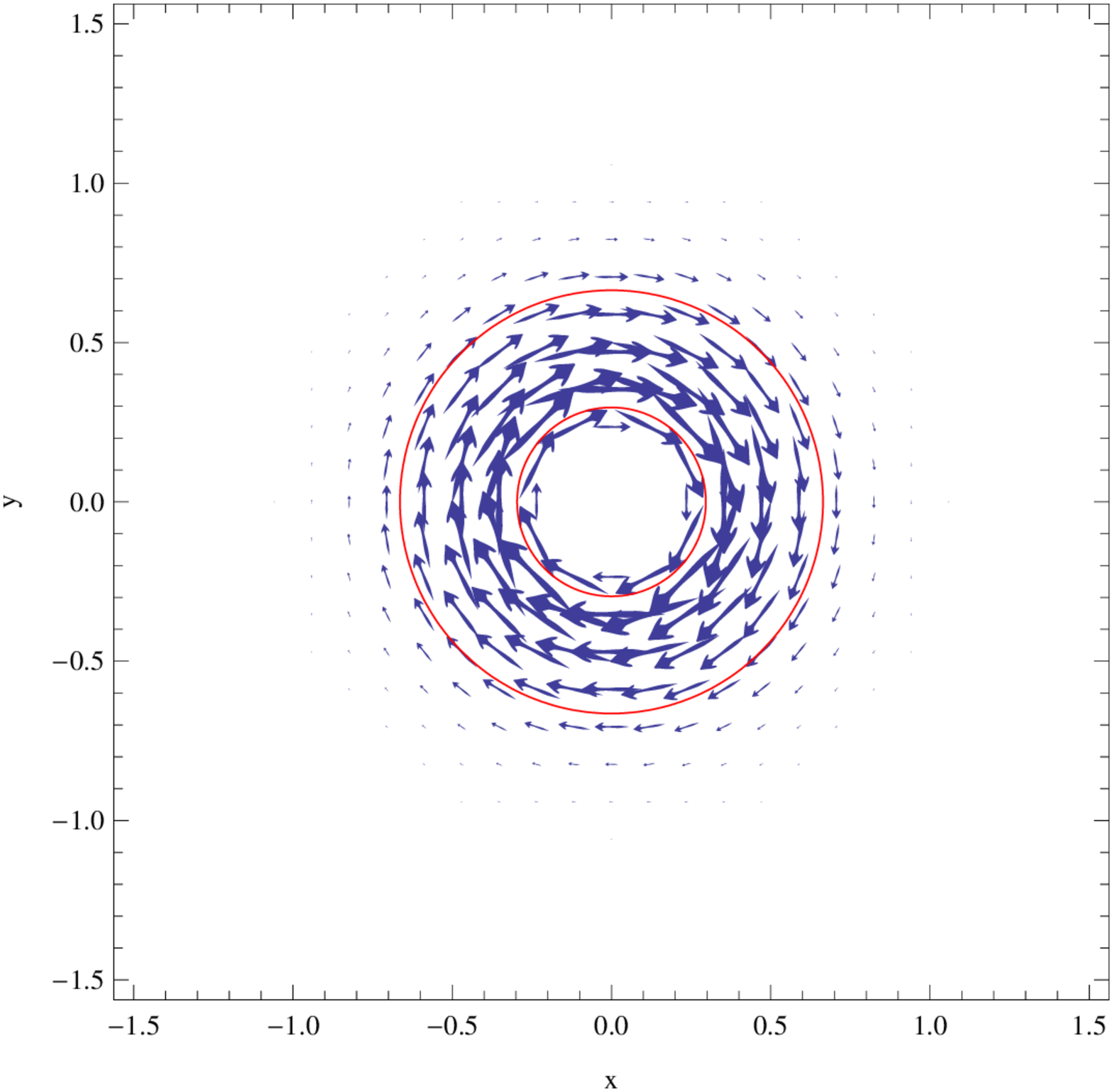}
\end{minipage}

\end{center}
\caption{\small Magnetic field orientation of the gauged
${\cal A}_{1,1}$ Hopfion. Top row: The magnetic flux in the $xz$ plane
at $g=0.1$ (left) and $g=2$ (right). The red oval profile indicates the position curve of the gauged
Hopfion given by the isosurface of the form $\phi^3 = -0.80$. Bottom row: The magnetic flux in the $xy$ plane
at $g=0.1$ (left) and $g=2$ (right). The position curve is indicated by the red curves.}
\end{figure}

It is instructive to compare our results with pattern of evolution of the gauged baby Skyrmions
\cite{Gladikowski:1995sc}. In the latter case the solitons also carry magnetic flux $\Phi = g \int_{\mathbb{R}^2} B$ which is
in general non-quantized. The flux of the gauged baby Skyrmions is associated with the position of the solitons, it
is orthogonal to the $xy$ plane. An interesting observation is that as gauge coupling grows,
the magnetic flux of the degree $n$ baby Skyrmions varies from 0 to $-2\pi n$, i.e. in the strong coupling regime
the magnetic flux is quantized thought there is no topological reasons for it.

Since the axially symmetric Hopfions can be thought as planar Skyrmions placed along a twisted closed
string \cite{Gladikowski:1996mb,Kobayashi:2013bqa}, this picture is certainly consistent with our results.
Indeed, as shown in  Fig.~\ref{fig:2}, the circular magnetic flux is orthogonal to the
$xz $ plane, although its radius is
slightly larger than the radius of the position curve of the Hopfion. Certainly, the total flux through the
$xz$ plane is zero, in order to evaluate the magnitude of the flux we have to consider the $xz$ half-plane or,
equivalently, the $z\rho$ plane.
We found that in the strong coupling limit
the magnetic flux of the gauged Hopfion through the $z\rho$ plane becomes quantized in units
of $2\pi$, see Fig.~\ref{fig:3}.

\begin{figure}[hbt]
\lbfig{fig:3}
\begin{center}
\includegraphics[height=.27\textheight, angle =0]{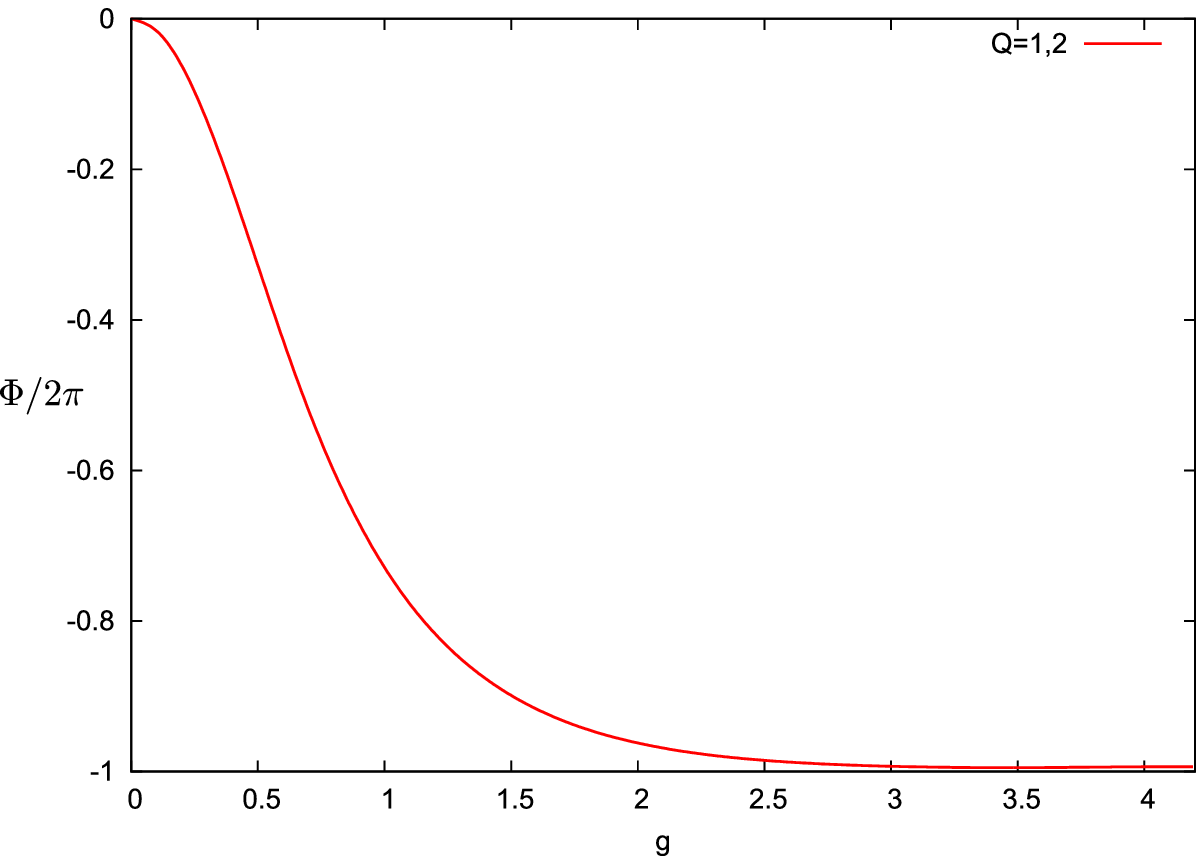}
\end{center}
\caption{\small The magnetic flux in units of $2\pi$ through the $z \rho$-plane as a function of the coupling constant $g$ for
for the solutions of degree $Q=1,2$. }
\end{figure}

A particularly interesting observation is that the total circular flux of the configurations of degrees $Q=1,2$
through the $z \rho$ plane shows the same dependence on the coupling constant, independently of the Hopf degree of the soliton.
We can understand this pattern when we recall that, as it was mentioned above, the axially
symmetric configuration of the type ${\cal A}_{m,n}$ can be thought of as composed from
planar baby Skyrmion of charge $n$ twisted $m$ times
along the circle. Thus, the circular magnetic flux is associated with the planar
charge $n=1$ which is the same in both cases we are considering.

The situation changes in the strong coupling limit $g  \geq g_{cr} \sim 3$. Then the contribution of the Maxwell
term becomes negligible and the condition of regularity of the energy functional
\re{E} is satisfied if $D_k\phi^a=0$ as $r \to \infty$. Furthermore,
our simulations show that at the
large coupling the component $A_\varphi(r,\theta)$ develops a sharp plateau $gA_\varphi = -m$ in the vicinity of the
position curve. Here the integer $m$, as defined in \re{ansatz}, corresponds to the number of twists along the position curve.
The plateau further extends as the gauge coupling grows.
On the other hand, in the strong coupling limit the position curve itself
expands from a circle $S^1$ to some region. In some sense it resembles the Meissner effect, at critical
value of the gauge coupling $g_{cr} \sim 3 $ the magnetic field is expelled from the Hopfion.

Indeed, in this region  the covariant derivative in azimuthal direction $D_\varphi \phi^a=0$
and the component $A_\varphi$ is a pure gauge, i.e.
$gA_\varphi  = - iU\partial_\varphi U^{-1}$, where $U=e^{i m\varphi}$. Clearly this corresponds to the
linear string of magnetic flux through the center of the Hopfion, which is
quantized in units of $2\pi$ and carries $m$ quanta.

We illustrated this observation in  Fig.~\ref{fig:5}, where both the profiles of the rescaled
component of the potential $A_\varphi$ in units of $g/m$
and the component $\phi^3$ in the $xy$ plane are shown in the weak and strong coupling regimes.

\begin{figure}[hbt]
\lbfig{fig:5}
\begin{center}
\includegraphics[height=.31\textheight, angle =-90]{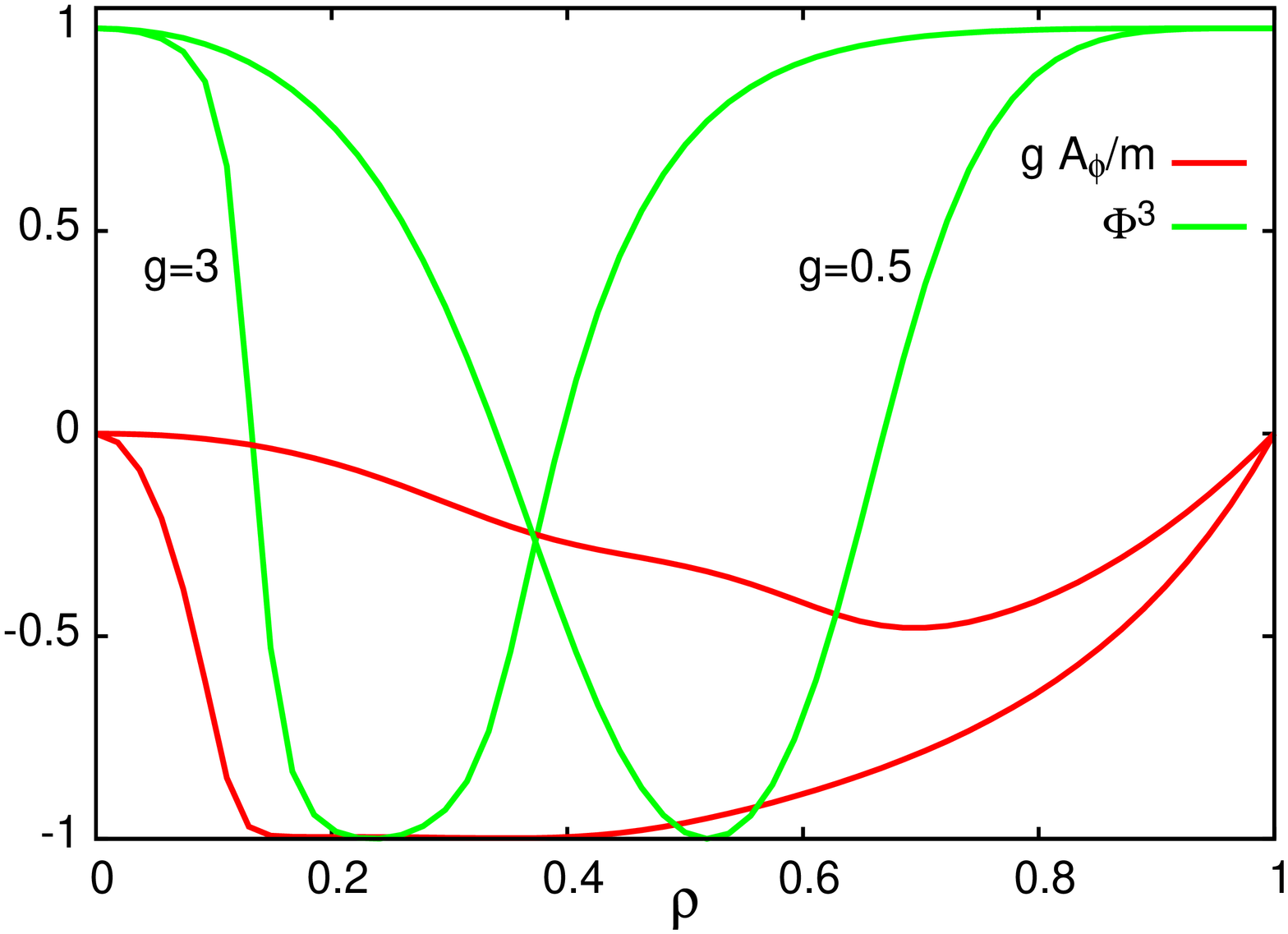}\hspace{0.1cm}
\includegraphics[height=.31\textheight, angle =-90]{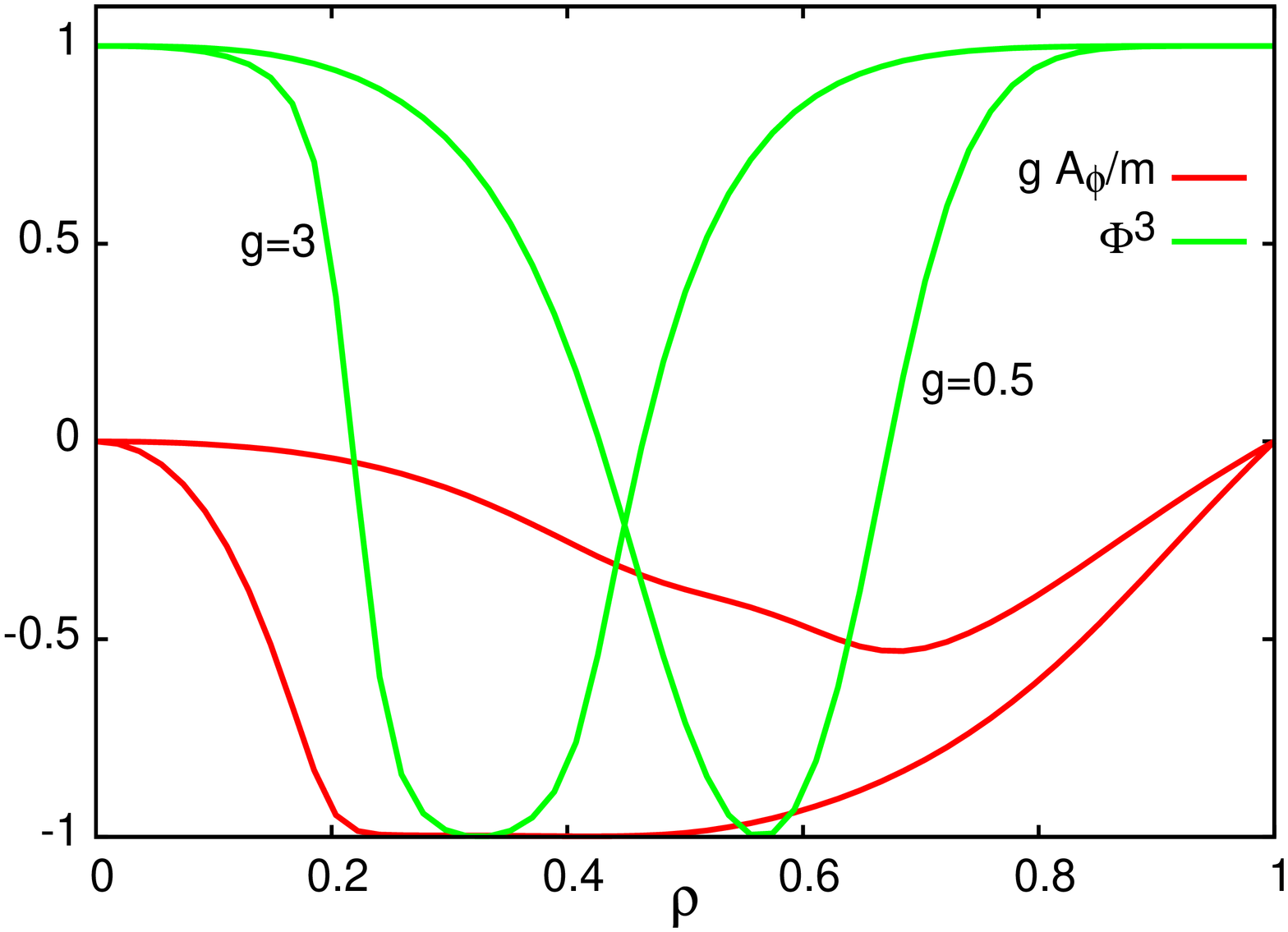}
\centerline{\hspace{0.5cm}${\cal A}_{1,1}$\hspace{7.5cm}${\cal A}_{2,1}$}
\end{center}
\caption{\small The magnetic flux $gA_\phi (\tilde x,0)/m$ (solid line) and
the scalar field component $\phi^3(\tilde x,0)$ (dashed line) of the gauged Hopfions of degree $Q=1,2$
in the $xy$ plane; the radial variable $\rho$ is compactified onto the unit interval, $\tilde x =
\rho/(1+\rho)\in$ $[0,1]$.}
\end{figure}

Hence, the corresponding integrated flux through the Hopfion along the symmetry axis is
 $-2\pi m$. Similar to the flux through the $xz$ plane, which encircles the Hopfion,
it becomes quantized in the strong coupling regime. We conclude that the gauged Hopfion carries \emph{two}
magnetic fluxes, the first circular flux encircles the position curve while the second one is directed along the
third axis. In the strong coupling regime they both are quantized in units of  $2\pi$ and  carry $n$ and $m$ quanta respectively.

Evidently, the appearance of the effective quantization of the magnetic field matches the underlaying topology of the Hopfion
configuration. This observation may be used to identify related topological invariant in the Maxwell sector of the gauged model.
This mechanism could allow us to reconsider the usual arguments concerning implementation of the
Protogenov-Verbus topological bound \cite{Protogenov:2002bt}. Physically, appearance of the quantized magnetic flux through the center
of the Hopfion may yield a constraint which could affect the usual scaling arguments \cite{Radu:2005jp}.


\section{Conclusions}

The main purpose of this work was to present a new type of gauged solitons in the Faddeev-Skyrme-Maxwell theory.
Our consideration is restricted to the simple axially-symmetric Hopfions ${\cal A}_{m,n}$ of lower degree $Q=mn$.
Similar to the corresponding solutions in the Skyrme model they are topologically stable, in the weak coupling regime
they carry non-integer toroidal magnetic flux. In the strong coupling regime the configuration is associated with
two  magnetic fluxes, one of which represent a circular vortex, and the second one is orthogonal to the position
curve. In this limit we observe an effective quantization of both fluxes, the first flux is quantized in units
of the winding number $n$ and the second flux is quantized in units of $m$, respctively.

Certainly, this is a first step towards complete investigation of the gauged Hopfions. Clearly, this study
should be extended to the Hopfions of higher degrees and different geometry.
One can expect by analogy with isorotations of the solitons of the Skyrme systems (see
\cite{BattyMareike,JHSS} and \cite{Halavanau:2013vsa,Battye:2013tka})
that the coupling of the Hopfions to the electromagnetic field may drastically affect their structure.
Since for a given degree $Q$ there are several different soliton solutions of rather
similar energy and the number of solutions seems to grow with $Q$ \cite{Sutcliffe:2007ui},
various bifurcations may occur as gauge coupling varies. Further, we cannot exclude possible
instabilities of gauged Hopfions at some critical coupling.

We do not investigate here  possible effects of the potential term in \re{model} on the properties of the
gauged Hopfions. We can expect, by analogy with the case of the gauged baby Skyrme model
\cite{Gladikowski:1995sc} that the results may strongly depend on ratio $\mu^2/g$. On the other hand
the limiting truncated Faddeev-Skyrme-Maxwell system which appears in the strong coupling limit should
be considered in depth, a possible interplay between the mass generating terms and the potential may drastically affect
the stability of the configuration.

An important feature of the Faddeev-Skyrme model is that the energy of the Hopfions is related with the
topological degree via the Vakulenko-Kapitanski bound  $E\ge Q^{3/4}$  \cite{VK}.
We expect that the coupling to the Maxwell field will affect this relation,
in particular, the solutions may approach the topological bound.
This problem will be investigated
elsewhere.

As a direction for future work, it would be interesting to study spinning gauged Hopfions,
these electrically charge solutions with an
intrinsic angular momentum can be constructed by analogy with similar configuration in the gauged Skyrme model
\cite{Radu:2005jp}.


\section*{Acknowledgements}
We thank David Foster, Olga Kichakova, Eugen Radu, Martin Speight and Tigran Tchrakian
for useful discussions and valuable comments.
This work is supported by the A.~von Humboldt Foundation in
the framework of the Institutes linkage Programm. We are
grateful to the Institute of Physics at the Carl von Ossietzky University
Oldenburg for hospitality.

\begin{small}

\end{small}

\end{document}